\documentclass[showpacs,preprintnumbers]{revtex4}
\usepackage{amsmath}
\usepackage{graphicx}

\begin{document}

\textbf{Comment on \textquotedblleft Prospects for a new search for the
electron }

\textbf{electric-dipole moment in solid gadolinium-iron-garnet
ceramics\textquotedblright }\bigskip

Tomislav Ivezi\'{c}

\textit{Ru%
\mbox
{\it{d}\hspace{-.15em}\rule[1.25ex]{.2em}{.04ex}\hspace{-.05em}}er Bo\v{s}%
kovi\'{c} Institute, P.O.B. 180, 10002 Zagreb, Croatia}

\textit{ivezic@irb.hr\bigskip }

In a recent paper [A. O. Sushkov, S. Eckel and S. K. Lamoreaux, Phys. Rev. A
\textbf{79}, 022118 (2009)] the authors measured the EDM-induced
magnetization $M$ that is given by Eq. (1) in their paper. Such an
expression for $M$ is a consequence of the generally accepted opinion that
both dipole moments, a MDM $\mathbf{m}$ and an EDM $\mathbf{d}$, are
proportional to the spin $\mathbf{S}$. Recently [T. Ivezi\'{c}, Phys. Scr.
\textbf{81}, 025001 (2010)] the Uhlenbeck-Goudsmit hypothesis is generalized
in a Lorentz covariant manner using the four-dimensional (4D) geometric
quantities. From the viewpoint of such formulation there is no EDM-induced
magnetization $M$; in the 4D spacetime the EDM $d^{a}$ is not proportional
to $S^{a}$. It is argued that the induced $M$ can come from the direct
interaction between the applied electric field $E^{a}$ and a MDM $m^{a}$%
.\bigskip \medskip

\noindent PACS number(s): 03.30.+p, 31.30.jp, 13.40.Em\medskip \bigskip

\begin{center}
\textbf{I. INTRODUCTION\bigskip }
\end{center}

In [1], the paramagnetic insulating sample is subjected to an electric field
$\mathbf{E}$, see the Introduction and Fig. 1 in [1]. It is supposed in [1],
as usual in the elementary particle theories, that not only the magnetic
dipole moment (MDM) $\mathbf{m}$ is proportional to the spin $\mathbf{S}$ ($%
\mathbf{m}=m(\mathbf{S}/S)$, the Uhlenbeck-Goudsmit hypothesis), but the
electric dipole moment (EDM) $\mathbf{d}$ as well (the usual 3-vectors are
written in boldface). The external electric field $\mathbf{E}$ orients
permanent EDMs along the field; the interaction term is $-\mathbf{d}\cdot
\mathbf{E}=-d(\mathbf{S}/S)\cdot \mathbf{E}$. Hence, the MDMs will be
oriented as well. It will cause that the sample acquires a net magnetization
$\mathbf{M}$ that is measured by a SQUID magnetometer, as the electric field
is reversed. The interaction between the electric field $\mathbf{E}$ and a
MDM $\mathbf{m}$ is only indirect through the alignment of three-dimensional
(3D) spins by the interaction $-d(\mathbf{S}/S)\cdot \mathbf{E}$. Thus, the
EDM-induced magnetization $\mathbf{M}$ is obtained%
\begin{equation}
M=\chi kd_{e}E/\mu _{a},  \label{ml1}
\end{equation}%
Eq. (1) in [1]; $M$ is determined by the electron EDM $\mathbf{d}_{e}$ and
the applied electric field $\mathbf{E}$. Hence, measuring the magnetization
the authors also indirectly measured the electron EDM.

Recently, [2], the Uhlenbeck-Goudsmit hypothesis is generalized in a Lorentz
covariant manner using 4D geometric quantities; the dipole moment tensor $%
D^{ab}$ is proportional to the spin four-tensor $S^{ab}$, $%
D^{ab}=g_{S}S^{ab} $, Eq. (9) in [2]. Using a general rule for the
decomposition of a second rank antisymmetric tensor, $D^{ab}$ is decomposed
according to Eq. (2) in [2]. The dipole moment vectors $d^{a}$ and $m^{a}$
are then derived from $D^{ab}$ and the velocity vector of the particle $%
u^{a} $ ($d^{a}$, $m^{a}$, $u^{a}$, $S^{a}$, etc. are usually called
4-vectors). Similarly, $S^{ab}$ is decomposed according to Eq. (8) in [2].
The usual \textquotedblleft space-space\textquotedblright\ intrinsic angular
momentum, spin $S^{a}$, and a new one, the \textquotedblleft
time-space\textquotedblright\ intrinsic angular momentum, spin $Z^{a}$, are
derived from $S^{ab}$ and $u^{a}$, Eq. (8) in [2]. Then, Eq. (10) in [2] is
obtained as%
\begin{equation}
m^{a}=cg_{S}S^{a},\ d^{a}=g_{S}Z^{a},  \label{dm}
\end{equation}%
According to (\ref{dm}), the intrinsic MDM $m^{a}$ of an elementary particle
is determined by $S^{a}$, whereas the intrinsic EDM $d^{a}$ is determined by
the new spin vector $Z^{a}$ and not, as usual, by the spin $\mathbf{S}$.
Both spins, $S^{a}$ and $Z^{a}$, are equally good physical quantities. The
EDM $d^{a}$, which is obtained in this way, i.e., from the connection with
the spin $Z^{a}$, Eq. (\ref{dm}), is an intrinsic property of elementary
particles in the same way as it is the MDM $m^{a}$. In contrast with it, in
the elementary particle theories, as mentioned in [2]: \textquotedblleft ..
an EDM is obtained by a dynamic calculation and it stems from an asymmetry
in the charge distribution inside a fundamental particle, which is thought
of as a charged cloud.\textquotedblright\ The EDM direction is connected
with a net displacement of charge along the spin axis, $\mathbf{d}=d\mathbf{S%
}/S$. The reason for the assumption that $\mathbf{d}$ has to be parallel to
the spin $\mathbf{S}$ comes from the general belief that $\mathbf{S}$ is the
only available 3-vector in the rest frame of the particle. However, as
noticed in [2]: \textquotedblleft ... neither the direction of $\mathbf{d}$
nor the direction of the spin $\mathbf{S}$ have a well-defined meaning in
the 4D spacetime. The only Lorentz-invariant condition on the directions of $%
d^{a}$ and $S^{a}$ in the 4D spacetime is $d^{a}u_{a}=S^{a}u_{a}=0$. This
condition does not say that $\mathbf{d}$ has to be parallel to the spin $%
\mathbf{S}$.\textquotedblright\ Obviously, the same remark holds if $\mathbf{%
d}$ is replaced by $\mathbf{m}$ and $d^{a}$ by $m^{a}$. More generally, from
the viewpoint of the geometric approach from [2], the 3D quantities $\mathbf{%
m}$, $\mathbf{d}$, $\mathbf{S}$, $\mathbf{E}$, $\mathbf{B}$, etc. are not
well-defined quantities in the 4D spacetime and they have to be replaced by
the 4D quantities, $m^{a}$, $d^{a}$, $S^{a}$, $E^{a}$, $B^{a}$, etc. The
results from [2] are in a complete agreement with the symmetry of the 4D
spacetime. They strongly indicate that the basic points of the
interpretation of measurements of EDM in [1], i.e., both $\mathbf{m}$ \emph{%
and} $\mathbf{d}$ are parallel to $\mathbf{S}$, are meaningless in the
manifestly covariant formulation from [2].

This means that \emph{in the Lorentz covariant formulation with 4D geometric
quantities it is not possible to infer anything about the electron EDM} $%
d_{e}^{a}$ \emph{from measurements of magnetization of a paramagnetic
insulating sample that is subjected to an electric field.} Instead of an
indirect interaction between the applied electric field $\mathbf{E}$ and a
magnetic dipole moment $\mathbf{m}$ through the alignment of 3D spins by the
interaction $-d(\mathbf{S}/S)\cdot \mathbf{E}$ we propose a direct, Lorentz
covariant, interaction between the applied electric field $E^{a}$ and a MDM $%
m^{a}$; the term in $L_{int}$ that is proportional to $E_{i}m_{k}$, Eq. (\ref%
{1}) below. This new, Lorentz covariant, interaction that can be used for
the interpretation of measurements of EDM in [1] will be exposed below using
the results from [2].\bigskip

\begin{center}
\textbf{II. LORENTZ COVARIANT\ INTERACTION\ BETWEEN\ }$F^{ab}$ \textbf{AND} $%
D^{ab}\bigskip $
\end{center}

The interaction between the electromagnetic field $F^{ab}$ and $D^{ab}$ is
given by the expression $(1/2)F_{ab}D^{ba}$, Eq. (3) in [2]. When the
decomposition of $F^{ab}$ (in terms of vector fields $E^{a}$, $B^{a}$ and
the velocity vector $v^{a}$ of the observers who measure fields), Eq. (1) in
[2], and the above mentioned decomposition of $D^{ab}$, Eq. (2) in [2], are
inserted into that expression then Eq. (3) in [2] is obtained. That equation
is first reported in [3].

As can be seen from the discussion of Eqs. (1) and (2) in [2], when it is
taken that the laboratory frame is the $e_{0}$-frame (the frame in which the
observers who measure $E^{a}$ and $B^{a}$ are at rest with the standard
basis $\{e_{\mu }\}$ in it), then the temporal components of $E^{a}$ and $%
B^{a}$ will be zero, $E^{0}=B^{0}=0$, and \emph{only} three spatial
components $E^{i}$ and $B^{i}$ will remain. Similarly, \emph{only} in the
particle's rest frame with the standard basis in it the dipole moments $%
d^{a} $ and $m^{a}$ will have $d^{0}=m^{0}=0$ and only three spatial
components $d^{i}$ and $m^{i}$ will remain. Thus, it is not possible that,
e.g., in the laboratory frame, \emph{both}, the fields and the dipole
moments have \emph{only} three spatial components, i.e., as for the usual
3-vectors. Thus, for example, in all EDM experiments the interaction between
the electromagnetic field and the dipole moments is described in terms of
the 3-vectors as $\mathbf{E}\cdot \mathbf{d}$ and $\mathbf{B}\cdot \mathbf{m}
$.

Furthermore, it can be seen from the discussion of Eq. (25) in [2] that in
the laboratory frame, as the $e_{0}$-frame, we can neglect the contributions
to $L_{int}$ from the terms with $d^{0}$ and $m^{0}$; they are $u^{2}/c^{2}$
of the usual terms $E_{i}d^{i}$ or $B_{i}m^{i}$. Then, what remains from Eq.
(3) in [2] is
\begin{equation}
L_{int}=-((E_{i}d^{i})+(B_{i}m^{i}))-(1/c^{2})\varepsilon
^{0ijk}(E_{i}m_{k}-c^{2}B_{i}d_{k})u_{j}.  \label{1}
\end{equation}%
This is, to order $0(u^{2}/c^{2})$, relativistically correct expression with
4D vectors for $L_{int}$. The last two terms that contain the direct
interactions between $E^{a}$ and $m^{a}$ and between $B^{a}$ and $d^{a}$ are
not taken into account in any of the EDM searches. With the usual 3-vectors,
it would correspond to Eq. (26) in [2]. But, as stated at the end of Sec. 5
in [2]: \textquotedblleft In the 4D geometric approach presented in this
paper the expressions like (26), (28) and (29) are meaningless, because, as
explained particularly in [12], there are not the usual time-dependent
3-vectors in the 4D spacetime.\textquotedblright\ Namely, [2]:
\textquotedblleft ... what is essential for the number of components of a
vector field is the number of variables on which that vector field depends,
i.e., the dimension of its domain. Thus, strictly speaking, the
time-dependent $\mathbf{E(r,}t\mathbf{)}$ and $\mathbf{B(r,}t\mathbf{)}$
cannot be the 3-vectors, since they are defined on the
spacetime.\textquotedblright\ Hence, contrary to the opinion of majority of
physicists, the usual formulation with 3-vectors $\mathbf{E}$, $\mathbf{B}$,
$\mathbf{S}$, etc. \textbf{IS NOT }relativistically correct formulation.

It is seen from the above Eq. (\ref{1}) for $L_{int}$ that the interaction
between the applied electric field and an EDM is contained in the term $%
-E_{i}d^{i}$. But, according to Eq. (\ref{dm}) $d^{a}$ is determined by the
\textquotedblleft time-space\textquotedblright\ spin $Z^{a}$, and not by the
3-vector spin $\mathbf{S}$. Furthermore, the interaction between the applied
electric field and a 4D magnetic moment $m^{a}$, which is determined by the
4D spin $S^{a}$, is contained in the term $-(1/c^{2})\varepsilon
^{0ijk}E_{i}m_{k}u_{j}$, which is $u^{a}$ - dependent. Hence, according to
this formulation, it is again visible that there is no EDM-induced
magnetization (the interaction term $-(1/c^{2})\varepsilon
^{0ijk}E_{i}m_{k}u_{j}$ does not contain $d^{a}$), but only EDM-induced
polarization (by means of $-E_{i}d^{i}$ term). This consideration also shows
that in the manifestly covariant formulation of the interaction between the
electric and magnetic fields and the dipole moments \emph{the magnetization
induced by the applied electric field can be explained only by the term} $%
-(1/c^{2})\varepsilon ^{0ijk}E_{i}m_{k}u_{j}$.

The same consideration can be completely applied to the recent
magnetization-based EDM search of the same authors that is presented in
[7].\bigskip

\begin{center}
\textbf{III. SOME\ ADDITIONAL\ REMARKS\bigskip }
\end{center}

It is declared by S. K. Lamoreaux, [4]: \textquotedblleft In the mean time,
I'll use the usual formulation of three vectors for the low velocities that
we use in EDM experiments. As the sources are not moving relative to the
boundaries of the experiment, this formulation is correct."

However, it has to be emphasized that the usual 3-vectors, e.g., $\mathbf{E}$
and $\mathbf{B}$ \textbf{ARE NOT} the low-velocity approximation of the 4D
vectors $E^{a}$ and $B^{a}$ and the usual transformations (UT) of $\mathbf{E}
$ and $\mathbf{B}$, Eq. (6) in [2], or Eqs. (11.148) and (11.149) from
Jackson's book (Ref. [10] in [2]), \textbf{ARE NOT} the low-velocity
approximation of the Lorentz transformations (LT) of the 4D vectors $E^{a}$
and $B^{a}$, Eq. (7) in [2]. Namely, according to the LT, e.g., the
components of the electric field 4D vector will be always transformed again
into the components of the electric field 4D vector; there is no mixing with
the components of the magnetic field 4D vector. It is just the opposite in
the UT. For more detail about the fundamental difference between the UT and
the LT of the electric and magnetic fields see, e.g., [5] or [6]. In [6], it
is shown that Minkowski first discovered the correct LT of the 4D electric
and magnetic fields, see also Ref. [12] in [2].

In addition, it is worthwhile to mention that in the approach from [2], see
the discussion in Sec. 5, neither the $T$ inversion nor the $P$ inversion
are good symmetries in the 4D spacetime, because they are synchronization
dependent.\bigskip

\begin{center}
\textbf{IV. CONCLUSIONS}
\end{center}

The consideration presented here shows that Eq. (1) from [1] (Eq. (\ref{ml1}%
) here), according to which the magnetization $M$ is determined by the
electron EDM $\mathbf{d}_{e}$, is not properly justified from the point of
view of the manifestly covariant formulation of the interaction between the
electromagnetic field and the dipole moments. It is argued that the
interaction term that is responsible for the induced magnetization by the
applied electric field in experiments in [1] is given by the third term in $%
L_{int}$, Eq. (\ref{1}), which does not contain the electron EDM $d^{a}$.

This means that, \emph{in order to get some useful informations about the
electron EDM in the experiments from }[1]\emph{\ (and }[7]\emph{), one would
need to measure a voltage induced by polarization due to the term} $%
-(E_{i}d^{i})$ \emph{in} $L_{int}$, \emph{Eq. }(\ref{1})\emph{, and not the
magnetization, because} $M$ \emph{cannot give any information about EDMs. }

Further examination of these results and their comparison with experiments
from, e.g., [1] (and [7]), requires much more work together with the
experimentalists who search for the electron EDM. Our aim was to explain
that the underlying physics (the 3D $\mathbf{m}$ \emph{and} $\mathbf{d}$ are
both parallel to $\mathbf{S}$) in experiments in [1] (and [7]), and in other
experimental searches for the electron EDM, is not well-founded in the 4D
spacetime.\bigskip

\noindent \textbf{References\bigskip }

\noindent \lbrack 1] A. O. Sushkov, S. Eckel, and S. K. Lamoreaux, Phys.
Rev. A \textbf{79}, 022118 (2009).

\noindent \lbrack 2] T. Ivezi\'{c}, Phys. Scr. \textbf{81}, 025001 (2010).

\noindent \lbrack 3] T. Ivezi\'{c}, Phys. Rev. Lett. \textbf{98}, 108901
(2007).

\noindent \lbrack 4] S. K. Lamoreaux, private communication.

\noindent \lbrack 5] T. Ivezi\'{c}, Found. Phys. Lett. \textbf{18}, 301
(2005).

\noindent \lbrack 6] T. Ivezi\'{c}, arXiv: 0906.3166v2

\noindent \lbrack 7] A. O. Sushkov, S. Eckel, and S. K. Lamoreaux, Phys.
Rev. A \textbf{81}, 022104 (2010).

\end{document}